Alessandro Pluchino, Cesare Garofalo, Giuseppe Inturri, Andrea Rapisarda, Matteo Ignaccolo

# Agent-based simulation of pedestrian behaviour in closed spaces: A Museum case study

 **Abstract**

In order to analyse the behaviour of pedestrians at the very fine scale, while moving along the streets, in open spaces or inside a building, simulation modelling becomes an essential tool. In these spatial environments, in the presence of unusual demand flows, simulation requires the ability to model the local dynamics of individual decision making and behaviour, which is strongly affected by the geometry, randomness, social preferences, local and collective behaviour of other individuals. The dynamics of people visiting and evacuating a museum offers an excellent case study along this line. In this paper we realize an agent-based simulation of the Castello Ursino museum in Catania (Italy), evaluating its carrying capacity in terms of both satisfaction of the visitors in regime of normal fruition and their safety under alarm conditions.

 **Motivation and Overview**

Walking is the most sustainable mode of transport. Survey data from a selection of seven European countries shows that 12-30% of all trips is made by walking (OECD, 1998). In Italy it involves 75% of all trips under a kilometer, as reported by ISFORT (2006), it is the first and last segment of every travel, affects the level of service of important transport infrastructure such as airports and railway stations. At the same time it is also of fundamental importance in fields related to urban planning, emergency, disaster planning. On the other hand, transportation engineering is traditionally focused on motorized travel and therefore there is a general lack of research and methods to model pedestrian behaviour. Existing transport pedestrian models can be roughly separated in analytical models and micro-simulations.

The first ones include "before and after" methods and regression analysis models (Older 1968, Pushkarev 1971), analogies with fluids, gas kinetics and other physical flow systems (Helbing 1992, Henderson 1974), entropy maximization (Butler 1978), dynamic network analysis with flow models calibrated on the basis of collected data (Di Gangi 2007), discrete choice models to predict pedestrians' route choice (Antonini 2006, Ignaccolo 2006), stochastic queuing and Markovian models (Mitchell 2001). In all these cases, the authors use mathematical models to calculate average pedestrian flows along a path, but these models are not able to include peculiar aspects of pedestrian (human) behaviour.

The second ones simulate the movement of each single pedestrian following a set of predetermined rules of behaviour and are applicable to a greater variety of situations, such as closed spatial environments or unusual demand flows, where local dynamics of individual decision making is strongly affected by geometry, randomness, social preferences, local and collective behaviour of other individuals. Helbing (1995) proposed a simulation approach based on the concept of "social force", that includes a sort of internal motivations of the individuals to perform certain actions (movements) and its influence on people's dynamic variables (velocity, acceleration, distance). Matrix-based systems like cellular au-tomata (CA) approach divide environments into cells and use cellular automata or similar methods to model movements of the pedestrians within cells. Blue (2001) used CA micro-simulation for modelling bi-directional pedestrian walkways, showing that a small rule set is capable of effectively capturing the behaviours of pedestrians at the micro-level while attaining realistic macro-level activity. Sarmady (2008) combined a behaviour model to simulate actions of individual pedestrians with a CA model used to simulate their small scale movements.

The overall output of these micro-level properties can lead to the emergence of a collective behavioural pattern, as the result of a trade-off between competitive and cooperative individual choices. The typical case is when local pedestrian movements towards some goal can lead to undesired crowded situations, while the tendency to follow what others are doing (herding effect) can favour congestion and panic. In other words, these simulation techniques are able to capture and explore rising crowd behaviours which cannot be described as simple aggregations of individual movements and that often are very far from intuition and hardly foreseeable by experience and common sense, as reported by Kitazawa (2004) and Osaragi (2004).

Batty (2001) argued that agent-based models have emerged as a powerful alternative to more aggregate and more geometric approaches to spatial modelling for many reasons: fine-scale data on urban land use, activities and flows of walkers are becoming more and more available; Chakrabarti et al. (2006) highlighted how computing capability continues to grow and new concepts on social complex systems are starting to blend with similar ideas in far-from-equilibrium physics. Agents simulations treat pedestrians as fully autonomous entities with cognitive and often learning capabilities (Koh, 2008; Batty, 1999). A comprehensive review of different approaches of computer simulation of evacuation can be found in Gwynne (1999). Based on a molecular-dynamic like micro-simulation approach, Helbing (2002) provided an overview on the observed collective phenomena in pedestrian crowds, such as lane formation in corridors and oscillations at bottlenecks in normal situations, while different kinds of blocked states are produced in panic situations.

In this paper we present an agent-based simulation of people moving in a closed spatial environment, considering the Castello Ursino museum in Catania (Italy) as a case study. The simulation intends to evaluate the so called "carrying capacity" of the museum, i.e. the total number of visitors the building can tolerate keeping maximum the level of global satisfaction (see also Garofalo, 2007) and taking into account the safety of the building, in terms of evacuation strategies and performances. In addition, the case study will be useful to show how an agent-based simulation of pedestrian walking dynamics in a given bounded environment can be important in exploring rising crowd behaviour, which can affect, in turn, the level of service and the safety of a closed environment.

##  The Castello Ursino museum case study

Using the software platform NetLogo (Wilensky, 1999), we apply the agent-based approach to study pedestrian behaviour inside the ground floor of a quite complex old building, the Castello Ursino, built in 1250 as one of the royal castles of Emperor Frederick II. It is a square compact building, with four towers and ten rooms placed around a central courtyard (Fig. 1), which in the last years has been used as a museum.

This building represents a good environment to test the complex emerging behaviour of a great number of visitors walking along its rooms full of paintings, sculptures and other works of art, trying to maximize their personal satisfaction during their visit experience, while interacting each other and with the environment. There is only one main entrance/exit door, but many rooms which contain different paintings and artworks, thus providing different levels of interest to visitors.

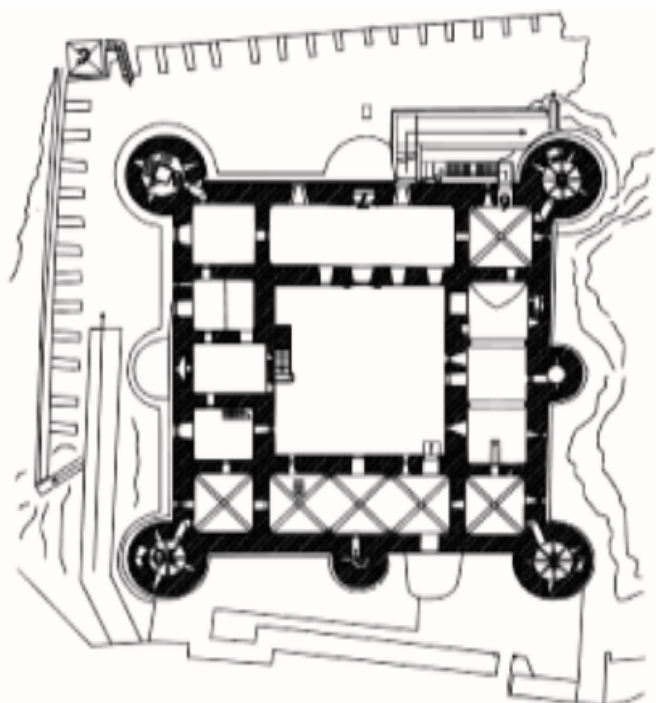

*Figure 1– Real Planimetry of the ground floor of Castello Ursino museum*

The aim of this study is to realize a virtual model of the museum in order to calculate its "carrying capacity". The concept of "carrying capacity" is commonly used in ecology and it is usually defined as the maximum population size of a given species that the environment can sustain indefinitely, given the food, habitat, water and other necessities available. Here, the same concept has been interpreted as the number of people, simultaneously visiting the museum, which maximizes their overall satisfaction under the safety conditions. We assume the latter as a suitable indicator of the performance of the virtual system under both the normal fruition dynamics and the emergency dynamics.

The fruition dynamics will be addressed in the next section, where the satisfaction of the visitors, together with their average visiting time, will be evaluated as function of the level of crowding within the museum. As most of the building belonging to historical heritage, Castello Ursino shows several concerns about safety management. Therefore the emergency dynamics will be also simulated, to measure the safety performances in terms of evacuation times for different layout of the buildings, safety procedures and crowding scenarios, as described in the last section. All the results will be averaged over several different simulation runs, performed with the same parameters.

## 🌐 Normal fruition dynamics: simulation results

The planimetry of the ground floor of Castello Ursino museum has been reproduced within the NetLogo interface preserving the real scale (Fig.2).

The grain of resolution is a square patch 60x60 cm$^2$, able to carry one visitor at a time. The different patch's colours of the spatial environment allow the agents to recognize and properly interact with objects with different "physical property", i.e. internal and external walls (in blue), external entrance/exit door (red on the bottom), internal doors (white), three emergency exits (green) from rooms 2, 4 and 8 to the internal courtyard (gray), one intermediate emergency exit (dark red) from the courtyard to room 10, artworks (yellow rectangular areas arbitrarily distributed inside the rooms), fruition spaces around the artworks (light yellow) and free space (in black). The green arrows in rooms 6, 7 and 10 represents emergency signs indicating the closer exit direction.

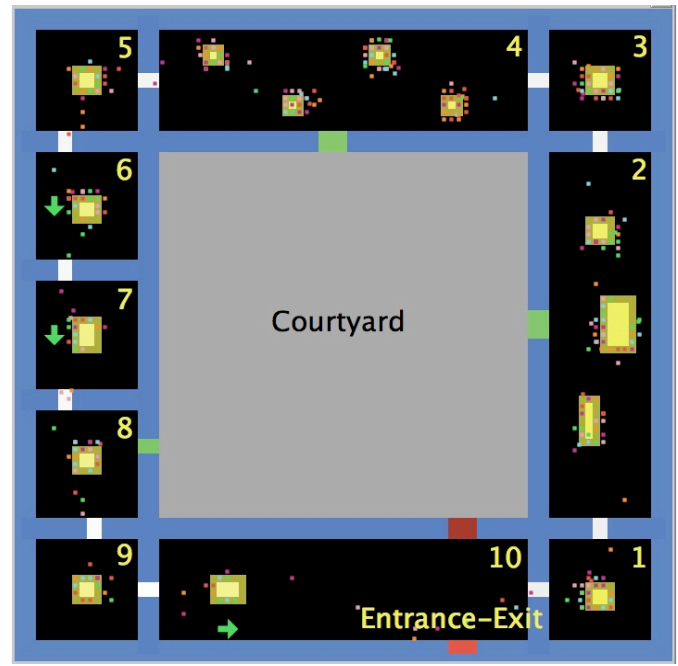

*Figure 2 – Virtual Planimetry of Castello Ursino, realized inside the NetLogo environment*

In the normal fruition dynamics, visitors (visualized with different colors in Fig.2) access the museum with a Poisson arrival, i.e. with exponentially distributed inter-arrival times $\Delta T$ (with a given mean $<\Delta T>$, expressed in seconds); they move counter-clockwise along one patch in one time-step (1 time step=1s), so they have an average velocity of 0.60 m/s, slightly less than the normal pace (0.75 m/s). The motion of the visitors is not predetermined and follows a few intuitive rules: each agent possesses a radius of vision and is attracted by doors or artworks around him; when no obstacles are present, he moves towards his target, doors or artworks, which are dynamically stored in an individual memory so that he will not pass twice through the same door and will not attend twice the same artwork; in presence of obstacles, walls or other people, he tries to avoid them going around in a random direction.

For each artwork, each agent has a different interest, i.e. the time he is willing to spend enjoying that artwork, and a different patience, i.e. the time he is willing to spend waiting for free space in the fruition area around that artwork before switching to the next artwork. These values are randomly extracted with uniform probability, for every visitor and for every artwork, within the intervals $[I_{min}=20s, I_{max}=50s]$ (for the interest) and $[P_{min}=30s, P_{max}=40s]$ (for the patience). These values are, of course, arbitrary, but plausible for our case study.

As a function of both the patience and the interest it is possible to calculate in real time the level of satisfaction of each visitor: in particular, individual satisfaction increases with the number of artworks really visited (proportionally to the interest) and decreases with the time spent in the fruition area for every artwork (proportionally to the patience), which in turn is affected by the presence of the other visitors. Once the visit is over, visitors exit from the only external door in room 10, which is the same used to access the museum, and their final satisfaction is added to a global variable, called 'total satisfaction'. The latter will be divided, in turn, by the number of visitors involved, in order to calculate our main variable, i.e. the 'average satisfaction'.

For evaluating the carrying capacity of the museum, we need to simulate the fruition dynamics for several stationary levels of crowding. These levels are obtained in correspondence of decreasing values of the mean inter-arrival time $<\Delta T>$ or, equivalently, for increasing values of the average flow of incoming visitors defined as $<F>=1/<\Delta T>$. For a given value of $<F>$, expressed here in visitors per minute (vis./m), we let the simulation run until a stationary number N$_{stat}$ of visitors are simultaneously present in the museum. In general this happens after a transient of around 2000 s.

In the top panel of Fig.3 we plot N$_{stat}$ as function of <F> and, as one could expect, we find that the stationary number of visitors inside the museum rapidly increases with the increasing of the incoming flow. Also the average visiting time <T$_{vis}$> (calculated over a sample of 200 agents in the stationary regime) increases with <F>, as shown in the middle panel of the same figure, but with a different slope: in fact, it stays quite constant for relatively small values of the incoming flow (up to about <F>=8 vis./min), then it starts increase; finally, above the threshold value <F> = 40 vis./min, it tends again to a stationary value (about 16 or 17 minutes) which no longer depends on the crowding.

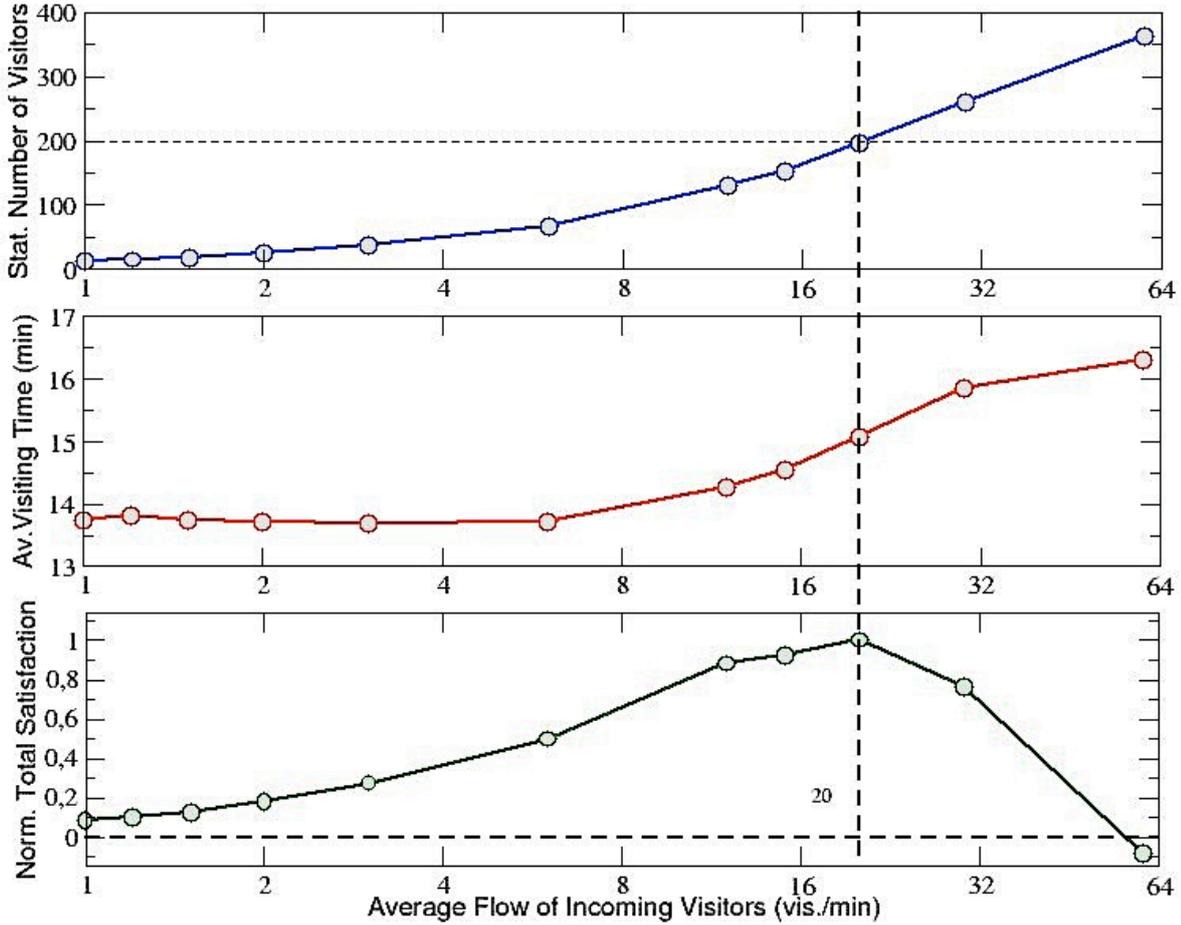

*Figure 3 – Top panel: The stationary number of visitors simultaneously present inside the museum is plotted for increasing values of the average flow of incoming visitors <F>. Middle panel: Average visiting time as function of <F>. Bottom panel: Normalized total satisfaction as function of <F>. An average over 5 events has been considered for each point of the three panels (the relative error for each point is of about 5%).*

Let us now evaluate how the value of the incoming flow affects the satisfaction of the visitors. If we call N$_{art}$ the total number of artworks in the museum (in our simulations N$_{art}$=15), the satisfaction of the *i*-th agent is defined as:

$$S(i) = \frac{\sum_{j=1}^{N_{art}} [I_j(i) - W_j(i)]}{S_{max}} \qquad (1)$$

where the $I_j(i)$'s are the (random) values of interest for the artworks visited by the i-th agent and the $W_j(i)$'s the waiting times the agent has spent in the fruition area of the same artworks. If for some artwork (say for j=k) the waiting time $W_k(i)$ overcomes the corresponding (random) patience $P_k(i)$, the agent will skip that artwork and we will put $I_k(i)=0$ and $W_k(i)=P_k(i)$ in Eq.(1). Finally, the normalization parameter $S_{max}$ is defined as $N_{art}$ ($I_{max} - P_{max}$). Of course S(i) grows or decreases as function of time until the visitor i-th is still inside the museum.

In Fig.4 we show the time evolution of both the real-time number of visitors inside the museum and their average real-time satisfaction <S> (calculated averaging Eq.(1) over all the visitors at time t), during the transient of 2000 s and for three events with increasing values of the average incoming flow (<F>=6, 20 and 60 vis./m respectively). The plots in the left column confirm what has been previously observed, i.e. that this transient is necessary to reach the stationary number $N_{stat}$ of visitors simultaneously present in the museum, a number that (as already shown in the top panel of Fig.3) increases with <F>. In the right column, the correspondent average satisfaction is reported as function of time: as one could expect, it tends to increase until the number of visitors also increases but, when the stationary number $N_{stat}$ is reached, it tends to an asymptotic value which decreases with <F>, becoming also negative for <F>=60 since each visitor has to share the fruition space with too many other visitors.

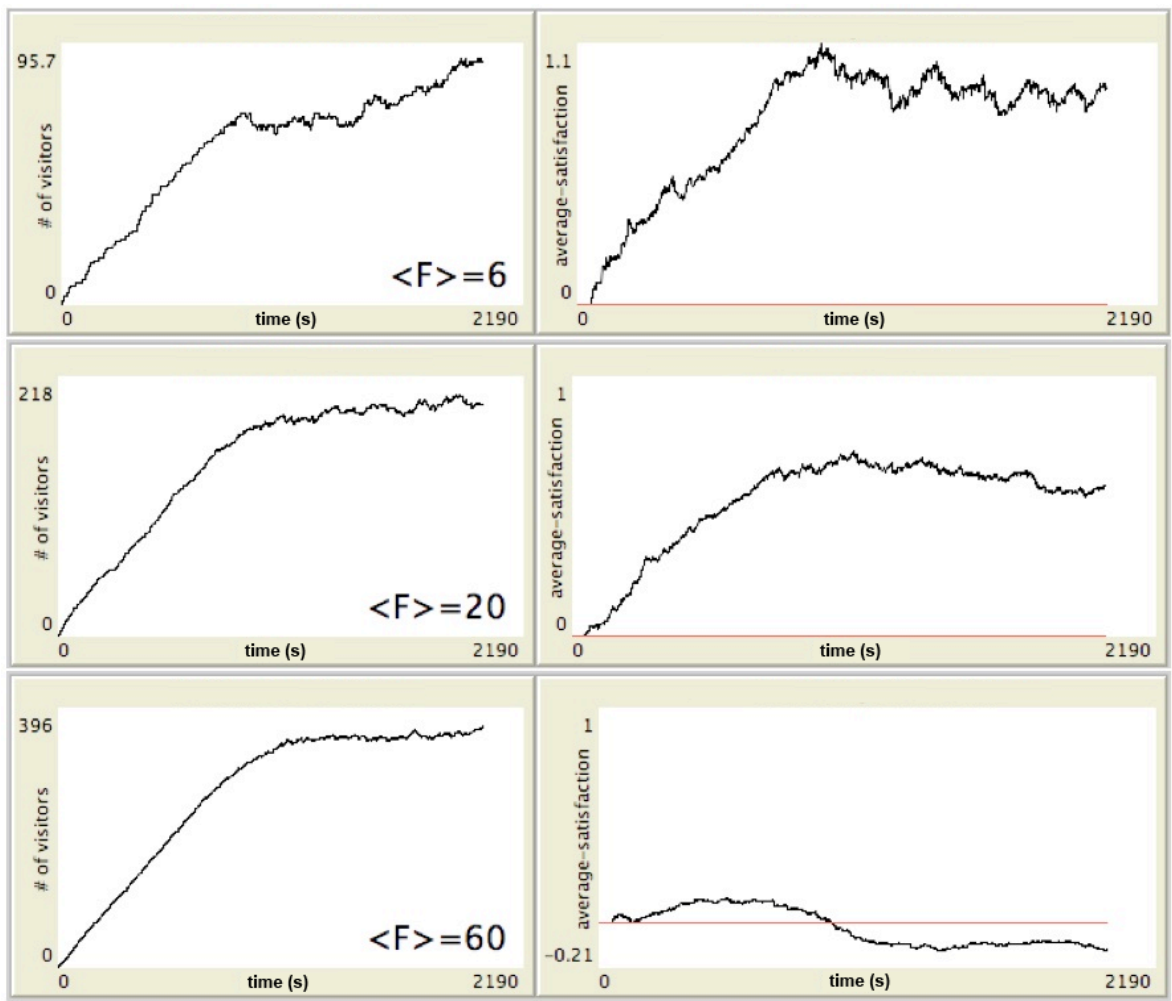

*Figure 4 – Real time evolution of both the number of visitors present inside the museum and their average satisfaction, for three single events with increasing values of the average incoming flow <F>.*

In order to evaluate the global level of satisfaction of all the visitors for increasing levels of stationary crowding, we now sum the satisfaction of each visitor leaving the museum *after* the transient of 2000 s. In particular, we perform this sum for samples of visitors with size equal to the value of $N_{stat}$ corresponding to the chosen value of <*F*> (following the top panel of Fig.3). In such a way we obtain the *total satisfaction* $S_{tot}$ (normalized to its maximum value and averaged over the usual 5 events) which is plotted, as function of <F>, in the bottom panel of Fig.3. Quite surprisingly, this plot presents a maximum that is reached for an average flow <F> of 20 visitors per minute, which in turn corresponds to a stationary number of visitors $N_{stat}=196$ (some dashed lines have also been plotted to guide the eye).

In order to better appreciate this interesting behaviour, in Fig.5 we plot again the total satisfaction, but now as a function of the stationary number of visitors simultaneously present inside the museum, $N_{stat}$. Furthermore, we report, in the same figure, also the average satisfaction per person (which corresponds to the asymptotic value shown in Fig.4, here averaged over 5 events) and we normalize the two quantities to their maximum values.

As we already observed in Fig.4, the average satisfaction per person (blue curve) decreases when the crowding increases. In particular, it presents its maximum value for the minimum load of the museum and tends to become negative as the number of visitors overcome 350 units.

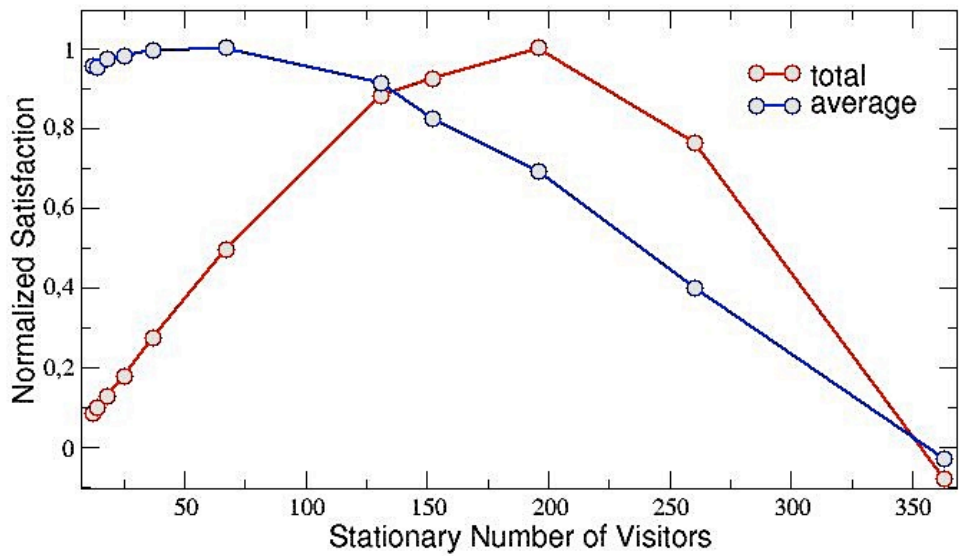

*Figure 5 – Normalized values of both the total and the average satisfaction as function of the stationary number $N_{stat}$ of visitors present inside the museum. The satisfaction is calculated over samples of $N_{stat}$ visitors who leave the museum and it is also averaged over 5 events.*

On the other hand, the behaviour of the total satisfaction of all the visitors (that can be also calculated as the product between the average satisfaction and the number of visitors) shows again the maximum in correspondence of about 200 agents observed in the bottom panel of Fig.3. We can conclude that such a value of crowding, evidently, represents the best compromise between the visitors' requirement of maximizing their personal satisfaction and the museum's requirement of maximizing the number of visitors simultaneously present inside the building.

On the basis of this result, that was very difficult to predict *a-priori*, we could reliably suppose that the number of 200 visitors is a good candidate to represent the real carrying capacity of the museum. Actually (see again the middle panel of Fig.3), this number results also compatible with a reasonable average visiting time (intermediate between the minimum and the maximum value) but we still have to test if it also allows to respect the safety conditions. In the next section we will check this requirement by simulating an emergency scenario.

 **Emergency dynamics: simulation results**

Let us focus on what happens when, after a transient regime of normal fruition of the museum, an alarm situation suddenly arises. In such a situation, of course, the behaviour dynamics of the agents changes as well as their average speed: in order to take into account the faster pace typical of an alarm situation, we rescaled the time by dividing it for a factor 1.66, thus obtaining an average velocity of 1m/s. We first consider the existing configuration of three internal emergency exits.

As it will be shown in the following, agent simulation allows a new approach, where configurational (number and location of exits), environmental (spatial geometry), behavioral (rules of agents) and procedural (emergency signs) aspects of the evacuation process have been simultaneously modeled, according to the scheme of Fig.6.

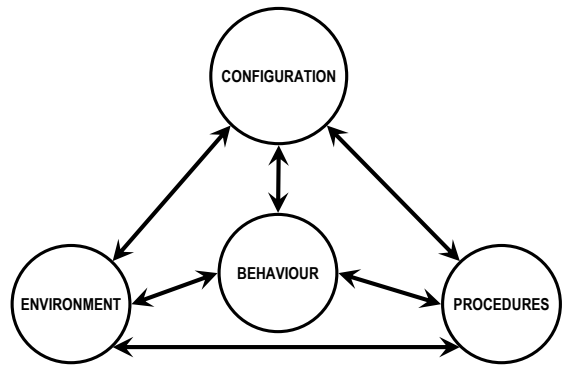

*Figure 6 – Scheme of the evacuation process*

Actually, as soon as an emergency alarm randomly goes off, each visitors checks if there is one of the three emergency exits overlooking the central courtyard within his radius of vision. If so, he moves towards that direction, otherwise, if emergency signs are present and visible to him, he follows them; if not, he goes backward following the same path where he came from, until the external entrance/exit door is found (see the flowchart in Fig.7). Of course, visitors gathered in the courtyard, rapidly make their way towards the intermediate emergency exit then, finally, they reach the exit door.

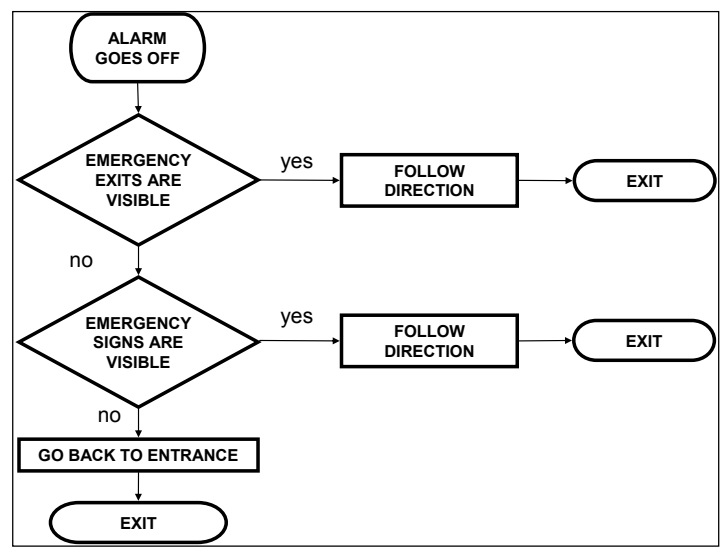

*Figure 7 – Flowchart of Emergency dynamics.*

A snapshot of a possible situation created by such a dynamics after an alarm event is shown, just to give an example, in Fig.8, where the configuration of three emergency exits (which is the existing one in the real museum) is reported.

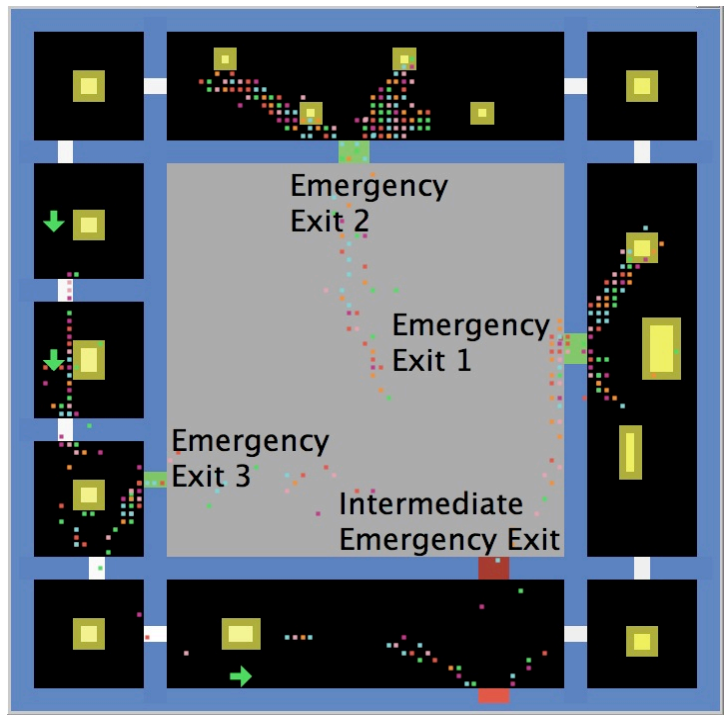

*Figure 8 – An example of evacuation through the courtyard in the existing configuration of three emergency exits.*

In our simulations of the emergency dynamics we let the alarm randomly go off after the usual transient of 2000 s, i.e. only after that, for a given average flow of incoming visitors $<F>$, the number $N_{stat}$ of visitors simultaneously present inside the museum has become stationary (as shown in the top panel of Fig.3, $N_{stat}$ is a function of $<F>$, therefore one may consider these two quantities as interchangeable). At this point, following the emergency flow chart of Fig.7, the visitors reach the nearest emergency exit within their radius of vision.

In Fig.9 we plot the number of visitors crossing the available emergency exits as function of time, for four increasing values of the incoming flow ($<F>=5,10,$ 20 and 60 vis./m). In particular, we consider three different configurations: the existing one (in the real museum), with three available emergency exits overlooking the courtyard, and other two alternative configurations with, respectively, only two (exit 1 and 2) and only one (exit 2) exits available.

As one can see, the majority of the visitors access to the courtyard through exit 2 (red curve), while only a small number of them use exit 3 (green curve). When the curves become horizontal (constant), it means that no more visitors are using the three exits, i.e. all visitors entered in the courtyard in order to reach the intermediate emergency exit. The time at which this happens visibly grows with the increasing flow (it goes from 40-50 s for $<F>=5$ to 120 s for $<F>=60$ vis./m, independently of the exit and of the number of exits available).

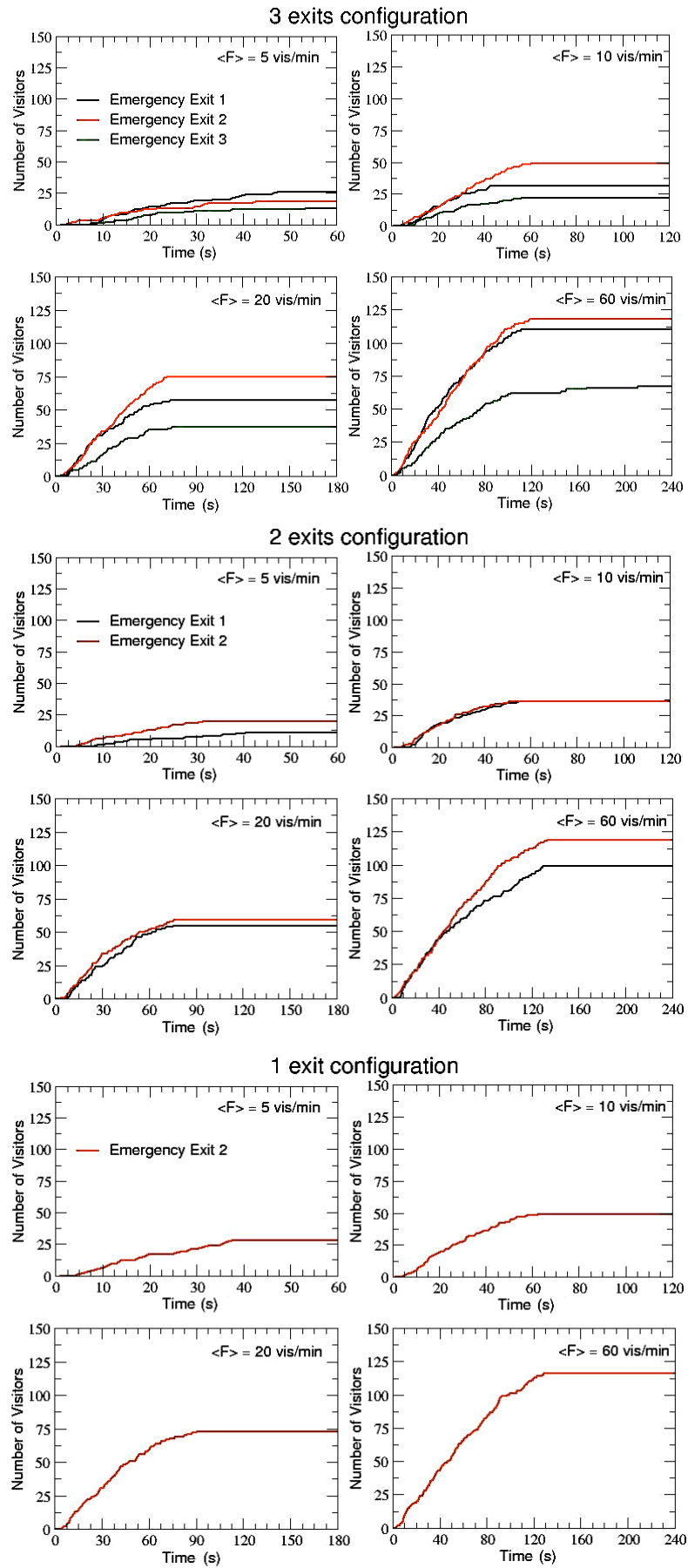

*Figure 9 – Emergency dynamics simulation. The number of visitors leaving the three emergency exits are reported as function of time for the three configurations with 3, 2 and 1 exits and, in each case, for increasing values of the average flow of incoming visitors.*

But not all the visitors use the three emergency exits overlooking the courtyard: actually, there is a certain number of visitors, in particular those situated in rooms 1 and 10 (see the planimetry in Fig.2), who does not see any emergency exit in their radius of vision, therefore – according to the dynamics described in the flowchart of Fig.7 – go directly to the main entrance/exit without passing through the courtyard. Such a component of visitors becomes evident if we plot the distribution of visitors, collected over five different independent simulations (events), as function of their respective escape time.

This is done in Fig.10, again for the three configurations of 3, 2 and 1 exit and for increasing values of <F>. In each panel are clearly distinguishable two peaks. The main one, on the right, includes the visitors using both the three emergency exits and the intermediate one, before reaching the main entrance/exit: these visitors need a longer time to evacuate, both for a longer distance to run (including the internal courtyard path) and for the crowding around the emergency exits. On the other hand, the lower peak on the left includes exactly that part of visitors closer to the main entrance/exit that rapidly evacuate the museum, going out with a shorter average escape time (from 10 s for <F>=5 vis./m to 40 s for <F>=60 vis./m).

As one could expect, a shift towards higher average escape times is visible for both the peaks when the incoming flow increases. This implies that the average escape time <$T_{esc}$>, calculated over the entire bimodal distribution and reported in the legend of each panel, also increases with <F> for the three configurations. In this respect, comparing <$T_{esc}$> between a configuration and the others for the same values of <F>, it is interesting to notice that <$T_{esc}$> stays quite constant or slightly oscillates for <F>=5 or 10 vis./min, while shows a decrease from 134 to 120 seconds passing from 3 to only 1 exit available: this seems to suggest that, from the point of view of the average escape time, it would be more effective to keep open only one emergency exit, i.e. the number 2, rather than two or three exits (see also Camillen et al., 2010).

A possible explanation of this apparently counterintuitive result could be related to the fact that, in order to minimize their own escape time, it is more convenient for the visitors situated in rooms 2, 6, 7, 8, 9 (in addition to those situated in rooms 1 and 10) when the alarm goes off, to run directly towards the main entrance/exit rather than to use the emergency exits 1 and 3.

But of course the average escape time could not be considered the best indicator for testing the effectiveness of the emergency system of the museum. More likely, one should be more interested in evaluating the global evacuation time, i.e. the time required for *all* the visitors to leave definitively the museum through the entrance/exit door of room 10.

In Fig.11 we plot such a quantity as function of the stationary number $N_{stat}$ of visitors present into the museum at the alarm moment (which, as previously observed, is in turn an increasing function of <F>), averaged over five different events. The three curves refer to the usual three exits configurations (with all the exits open, with exits 1 and 2 open and with only exit 2 open, respectively). Again, as one could expect, the global evacuation time increases increasing $N_{stat}$ for all the three configurations.

Notice that the resulting graph can be roughly divided in two different parts: in the first part, for $0 < N_{stat} < 215$, the 3-exits configuration curve (green) stays always below the other two curves, thus resulting more convenient; in the second part, for $N_{stat} > 215$, it is the 2-exits configuration that seems more effective, while – on the other end – the 1-exit configuration curve diverges for high levels of crowd. But just around an average number of 215 visitors, the three configurations appear to behave in the same way, giving an average evacuation time of 270 seconds, a little less than 5 minutes. Quite interestingly, this number of visitors is very close to 200, i.e. to the optimal number that, as we shown in the previous section, would represent the carrying capacity of the museum in terms of collective satisfaction.

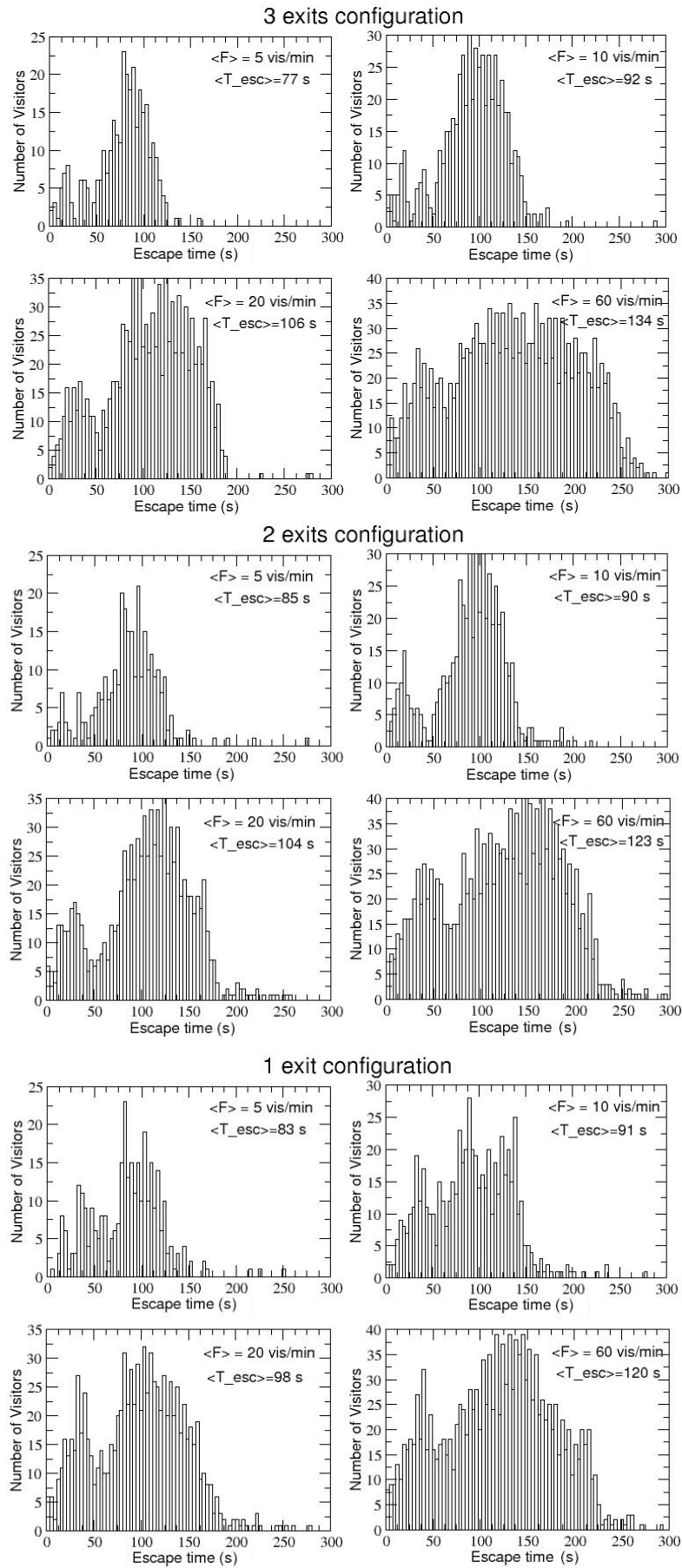

Figure 10 – Emergency dynamics simulation. The frequency distributions of the visitors as function of the single agents' escape time are reported for the three configurations with 3, 2 and 1 exit and, in each case, for increasing values of the average flow of incoming visitors.

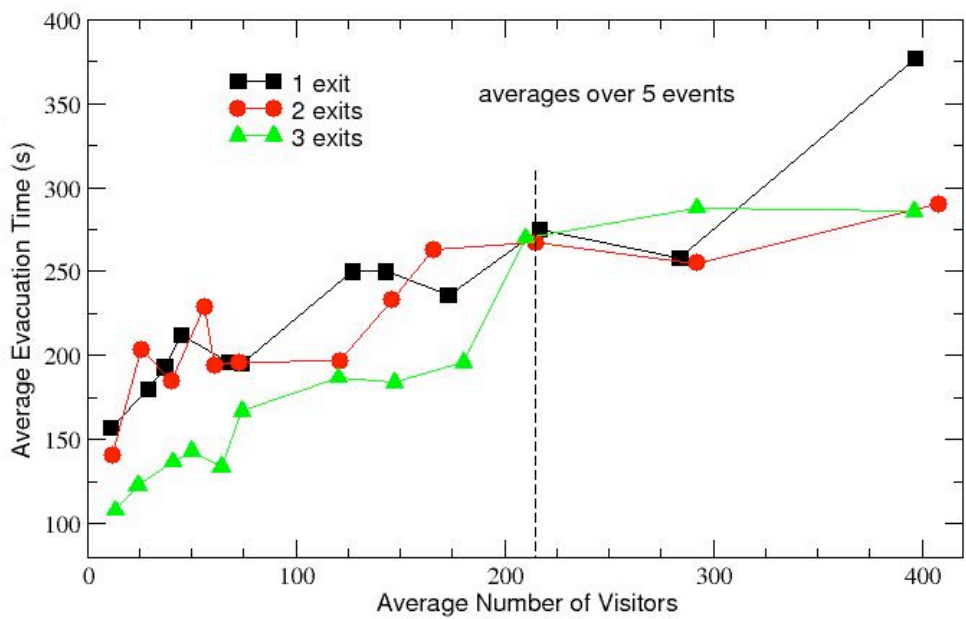

*Figure 11 – Emergency dynamics simulation. The global evacuation time, averaged over 5 different events, is reported as function of the (average) stationary number of visitors $N_{stat}$ inside the museum when the alarm goes off.*

We can conclude that the global maximum of satisfaction is perfectly compatible with a good level of safety, i.e. with a reasonable evacuation time, independently from the configuration of emergency exits adopted. Therefore, since the number of 200 visitors represents the best compromise between satisfaction and safety constraints, it can be reliably considered as the best carrying capacity estimate of the Castello Ursino museum.

## Conclusions

Agent-based simulations show their potential in many context of transport management in presence of unusual demand. We illustrated these ideas with an example based on the simulation of people visiting and evacuating a museum. This case study offers an excellent test environment for simulating a collective behavior emerging from local movements in a closed space. In the normal fruition regime we found the optimal number of visitors ensuring a maximum in the global satisfaction, while in the alarm regime we tested the effectiveness of the existing emergency plans for different exits configurations.

Merging the results of all these simulations, we confidently fixed the best carrying capacity estimate of the museum around 200 visitors simultaneously present inside. These findings confirm the convenience and the effectiveness of agent-based simulations in the design and analysis of complex social systems, fully supporting more traditional strategies already available to the engineers. They show their potential in many context of transport management in the presence of unusual demand and in situations where direct observations of human behaviour are not possible or not ethical, such as panic and emergency.

Agent-based simulations can be easily extended to other pedestrian "transportation environments", such as airports, ships, railway stations or urban areas. In this respect, several tasks can be performed, such as analysis of pedestrian safety, comfort and walkability, analysis of congestions, bottlenecks and level of service, feasibility studies of urban spaces.